\documentclass[12pt]{article}

\usepackage{epsfig}

\def\Journal#1#2#3#4{{#1} {\bf #2} (#4) #3}

\def\NPB{{\em Nucl. Phys.} B}
\def\PLB{{\em Phys. Lett.} B} 
\def\PRL{\em Phys. Rev. Lett.} 
\def\PRD{{\em Phys. Rev.} D}

\def\SPU{\em Sov. Phys. Usp.}
\def\PRC{{\em Phys. Rep.} C}
\def\IJMPA{{\em Int. J. Mod. Phys.} A}

\newcommand{\lwig}{\mbox{\,\raisebox{.3ex}
    {$<$}$\!\!\!\!\!$\raisebox{-.9ex}{$\sim$}\,}}

\newcommand{\iai}{I\overline{I}}

\newcommand{\ii}{{\rm i}} 
 
\newcommand{\xpr}{{x^\prime}}
 
\newcommand{\xbj}{x_{\rm Bj}}
\newcommand{\ybj}{y_{\rm Bj}}

\date{}

\begin{document}
\title{{\normalsize\rightline{DESY 96-202}\rightline{hep-ph/9609445}} 
  \vskip 1cm 
  \bf Instantons in Deep--Inelastic Scattering\\ -- The Simplest Process --
  \vspace{11mm}}
\author{S. Moch, A. Ringwald and F. Schrempp\\[12pt] 
Deutsches Elektronen-Synchrotron DESY, Hamburg, Germany}
\begin{titlepage} 
\maketitle
\begin{abstract}
  Instanton calculations in QCD are generically
  plagued by infra\-red  divergencies associated with the integration
  over the instanton size $\rho$. Here, we demonstrate explicitly 
  that the typical inverse hard momentum scale ${\mathcal Q}^{-1}$ 
  in deep-inelastic scattering provides a dynamical infrared cutoff for the
  size parameter $\rho$.
  Hence, deep-inelastic scattering may be viewed as a distinguished
  process for studying manifestations of QCD-instantons. For clarity,
  we restrict the explicit discussion to  the simplest chirality-violating 
  process,
  $\gamma^{\ast}+{\rm g}\rightarrow \overline{{\rm q}_{L}}+{\rm q}_{R}\,$. 
  We calculate the corresponding fixed angle cross-section as well as 
  the contributions to the gluon structure functions, ${\mathcal F}_{2\,g}$ 
  and ${\mathcal F}_{L\,g}$, within standard instanton perturbation theory 
  in leading semi-classical approximation. To this approximation, 
  fixed-angle scattering processes at high $Q^2$ are reliably 
  calculable. In the Bjorken limit, the considered instanton-induced 
  process gives a scaling contribution to ${\mathcal F}_{2\,g}(x,Q^{2})$ and
  the Callan-Gross relation holds. 
  \end{abstract} 
\thispagestyle{empty}
\end{titlepage}
\newpage \setcounter{page}{2}

\section{Introduction}

Instantons~\cite{bpst} are well known to represent tunnelling transitions in
non-abelian gauge theories between degenerate vacua of different topology. 
These transitions induce processes which are {\it forbidden} in perturbation
theory, but have to exist in general~\cite{th} due to Adler-Bell-Jackiw
anomalies. Correspondingly, these processes imply  a violation of 
certain fermionic quantum numbers, notably, $B+L$ in the electro-weak gauge
theory and chirality ($Q_5$) in (massless) QCD.
  
An experimental discovery of such a novel, non-perturbative
manifestation of non-abelian gauge theories would clearly be of basic
significance.

A number of results has revived the interest in instanton-induced processes
during recent years: 
\begin{itemize}
\item First of all, it was shown~\cite{r,m} that the generic exponential 
suppression of these tunnelling rates, $\propto \exp (-4\pi /\alpha )$, may be
overcome at {\it high energies}, mainly due to multi-gauge boson emission
in addition to the minimally required fermionic final state.
 
\item A pioneering and encouraging theoretical estimate 
of the size of the instanton ($I$) induced contribution to the 
gluon structure functions in deep-inelastic scattering  
was recently presented in Ref.~\cite{bb}. The summation over the $I$-induced 
multi-particle final state was implicitly performed by 
starting from the optical theorem
for the virtual $\gamma^\ast g\rightarrow \gamma^\ast g $ 
forward amplitude. The strategy was then to
evaluate the contribution to the functional integral coming from the
vicinity of the instanton-antiinstanton ($\iai$) configuration in 
Euclidean space, to analytically 
continue the result to Minkowski space and, finally, to take the imaginary
part. While the instanton-induced contribution to the 
gluon structure functions turned 
out to be small at larger values of the Bjorken variable $x$,
it was found in Ref.~\cite{bb} to increase dramatically towards 
smaller $x$.
\item Last not least, a systematic phenomenological and theoretical study
is under way~\cite{rs,grs,rs1,ggmrs}, which  clearly indicates that 
deep-inelastic $e p$ scattering at HERA now offers a unique window to 
experimentally detect QCD-instanton induced processes through their 
characteristic final-state signature. The searches for instanton-induced 
events have just started at HERA and a first upper limit of $0.9$ nb 
at $95\%$ confidence level for the cross-section of QCD-instanton induced 
events has been placed by the H1 Collaboration~\cite{H1}. New, improved 
search strategies are being developped~\cite{ggmrs} with the help of a 
Monte Carlo generator (QCDINS 1.3)~\cite{grs} for instanton-induced events.
\end{itemize}

The central question is, of course, whether instanton-induced processes
in deep-inelastic scattering can both be reliably computed and
experimentally measured. In particular, whether contributions associated
with the non-perturbative vacuum 
structure can be controlled in the same way as perturbative short-distance
corrections, in terms of a hard scale ${\mathcal Q}$.

In the work of Refs.~\cite{bb,bbgg} on deep-inelastic scattering,
the integrals over the instanton size $\rho$ were found to
be infrared (IR) divergent, like in a number of previous 
instanton calculations in different areas. Yet, the authors claimed  
that this problem does not affect the possibility, to isolate in 
deep-inelastic scattering a well-defined, IR-finite and sizable
instanton contribution in the regime of small QCD-gauge coupling, on 
account of the (large) photon virtuality $Q^2$. The IR-divergent pieces of
the $I$-size integrals were supposed to be factorizable into the parton
distributions, which anyway have to be extracted from experiment at 
some reference scale.       

On the other hand, also IR-finite instanton contributions to certain
observables in momentum space have been found in the past~\cite{as,early}. 
In this ideal case, the size of the contributing 
instantons is limited  by the inverse momentum scale ${\mathcal Q}^{-1}$ of 
the experimental probe, as one might intuitively expect. No {\it ad hoc} 
cutoff or assumption about the behaviour of large, overlapping instantons
need be introduced. 

A main issue of the present work is to shed further light on these important 
questions around the IR-behaviour associated with the instanton size in
deep-inelastic scattering.
This paper represents the first of several 
papers in preparation~\cite{mrs}, containing our theoretical results on  
$I$-induced processes in the deep-inelastic regime.

%%%%%%%%%%%%%%%%%%%%%%%%%%%%%FIGURE  %%%%%%%%%%%%%%%%%%%%%%%%%%%%%%%%%%%
\begin{figure} 
\begin{center}
  \epsfig{file=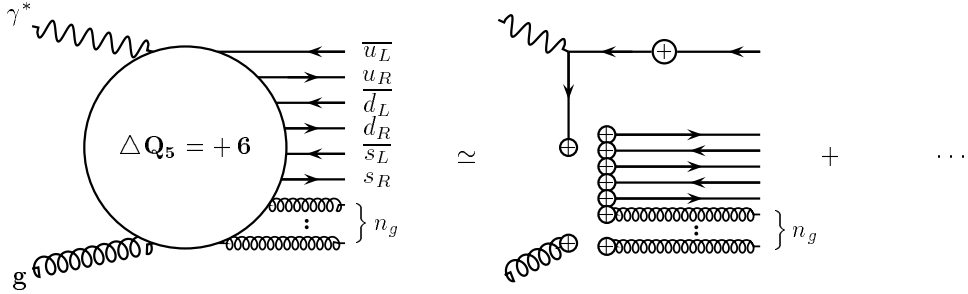,width=13cm}
\caption[dum]{\label{f1}
  Instanton-induced {\it chirality-violating} process,\\ $\gamma^{\ast}
  +{\rm g}\rightarrow \sum_{\rm flavours}^{n_{f}}\left[ \overline{{\rm 
      q}_{L}} +{\rm q}_{R}\right] +n_{g}\,{\rm g}$, corresponding to 
   three massless flavours 
  ($n_{f}=3$).}
 \end{center}
\end{figure}
%%%%%%%%%%%%%%%%%%%%%%%%%%%%%%%%%%%%%%%%%%%%%%%%%%%%%%%%%%%%%%%%%%%%%% 

For clarity, let us reduce here the realistic task of evaluating 
the $I$-induced cross-sections of the chirality violating multi-particle 
processes (illustrated in Fig.~\ref{f1})
\begin{equation}
  \gamma^{\ast}+{\rm g}\Rightarrow 
   \overline{{\rm u}_{L}}+{\rm u}_{R}\, 
   + \overline{{\rm d}_{L}}+{\rm d}_{R}\,
    +\overline{{\rm s}_{L}}+{\rm s}_{R}\,          
     +n_g\, {\rm g},
\end{equation}
to the detailed study of the {\it simplest} one,
without additional gluons and with just one massless flavour ($n_f=1$), 
\begin{equation}
  \gamma^{\ast}(q)+{\rm g}(p)\Rightarrow 
   \overline{{\rm q}_{L}}(k_{1})+{\rm
    q}_{R}(k_{2})\, .
\end{equation}
The price is, of course, that this process only represents a small fraction of
the total $I$-induced contribution to the gluon structure functions. 

However, there are also a number of important virtues: 

The present calculation provides a clean and explicit discussion of most of
the crucial steps involved in our subsequent task~\cite{mrs} to calculate 
the {\it dominant} $I$-induced contributions. While unessential technical
complications have been eliminated here, a generalization
to the realistic case with gluons and more flavours in the final state is
entirely straightforward~\cite{mrs,q96}.  
We shall explicitly calculate the corresponding fixed angle cross-section 
and the contributions to the gluon structure functions in leading
semi-classical approximation within standard instanton perturbation theory
(Sect.~3). Gauge invariance is kept manifest along the 
calculation and we may compare at various stages with the appropriate 
chirality-conserving process, calculated in leading order of perturbative QCD. 

As a central result of this paper and unlike Ref.~\cite{bb}, we find 
{\it no} IR divergencies associated with the integration over the 
instanton size $\rho$, which can even be perfomed analytically.  We are able 
to demonstrate explicitly that the typical hard scale ${\mathcal Q}$ 
in deep-inelastic scattering provides a {\it dynamical} infrared cutoff for 
the instanton size, $\rho \lwig {\mathcal O}(1/{\mathcal Q})$. 
Additional gluons in the final state will not change this conclusion, as is
briefly outlined in Sect.~4. Thus, deep-inelastic
scattering may indeed be viewed as a distinguished process for studying 
manifestations of QCD-instantons.

\section{Setting the Stage}
 
Let us start with the matrix elements 
${\mathcal T}^\mu (q,p; k_{1},\ldots ,k_{n})$ of the 
general, exclusive photon-parton reactions
\begin{equation}
  \gamma^{\ast}(q)+ p\rightarrow k_{1}+\ldots +k_{n}\, ,
\end{equation} 
in terms of which we form the inclusive structure tensor
${\mathcal W}^{\mu \nu}_{p}$ of the parton $p$,
\begin{eqnarray}
  {\mathcal W}^{\mu \nu}_{ p} (q,p ) & =&  \sum_{n=1}^\infty
  {\mathcal W}^{\mu \nu\, (n)}_{ p} (q,p ) \, ,
\label{wmunu}
\hspace{24pt} {\rm with} \\[10pt] 
 {\mathcal W}^{\mu \nu (n)}_{ p} ( q,p )& =& 
\frac{1}{4\,\pi}\,
  \int dPS^{(n)}\, 
    {\mathcal T}^\mu (q,p;k_{1},\ldots,k_{n}) 
    {\mathcal T}^{\nu\,\ast}(q,p;k_{1},\ldots ,k_{n} )  ,
\label{wmunu_n}
\end{eqnarray}
and 
%\hspace{12pt} 
%{\rm and} \\[10pt] 
\begin{eqnarray}
\int dPS^{(n)} = \prod_{j=1}^{n} \int
\frac{d^{4}k_{j}}{(2\,\pi)^{3}}\, \delta^{(+)}\left( k_{j}^{2}\right)
(2\,\pi)^{4}\, \delta^{(4)}\left( q+p-k_{1}-\ldots -k_{n}\right) \, .
\end{eqnarray}
Averaging over colour and spin of the initial state is implicitly understood
in Eq.~(\ref{wmunu_n}); the index $n$ is to label besides the final
state partons also their spin and colour degrees of freedom.

For exclusive $2\rightarrow 2$ processes, 
$\gamma^{\ast}(q) +p\rightarrow k_{1}+k_{2}$, the differential cross-section
is then expressed as 
\begin{equation}
\label{diffcross}
\frac{d\sigma}{dt}= 8\,\pi^{2}\,\alpha\ \frac{x}{Q^{2}}\,
\left[  -g_{\mu \nu}\, \frac{d{\mathcal W}^{\mu \nu\, (2)}_{p}}{dt}
\right] \, ,
\end{equation}
where  $Q^{2}=-q^{2}$ denotes the photon virtuality and
\begin{equation}
\label{t}
t=\left(  q-k_{1}\right)^{2}=\left(  p-k_{2}\right)^{2}  
\, ,
\end{equation}
the momentum transfer squared.

In general, each final state, $k_{1}+\ldots +k_{n}$, contributes
to the structure functions ${\mathcal F}_{i\,{ p}}$ 
of the parton $p$ via the projections 
\begin{eqnarray}
  {\mathcal F}^{(n)}_{2\,{ p}} (x ,Q^2) &= & \left[- g_{\nu\mu }\,
  +6\,x\,\frac{p_\mu p_\nu}{p\cdot q} \right]
   x\,{\mathcal W}_{ p}^{\mu \nu\, (n)} (q,p) \,
  ,
\label{projF2}
\\[10pt] 
{\mathcal F}^{(n)}_{L\,{ p}} (x ,Q^2) 
&= & 4\,x^2\,\frac{p_\mu p_\nu}{p\cdot q} 
{\mathcal W}_{ p}^{\mu \nu\, (n)} (q,p) \ ,
\label{projFL}
\hspace{24pt} {\rm such\ that} \\[10pt]
{\mathcal F}_{i\,{ p}}(x,Q^{2})&=&\sum_{n=1}^{\infty}
{\mathcal F}^{(n)}_{i\,{ p}}(x,Q^{2})\, ,
\end{eqnarray}
with
\begin{equation}
  x\equiv \frac{Q^2}{2\,p\cdot q} \, .
\end{equation}
denoting the Bjorken variable of the photon-parton subprocess.

The spin averaged proton structure functions $F_2$ and $F_L$, appearing
(in the one photon exchange approximation) in the unpolarized inclusive
lepto-production cross-section as
\begin{equation}
   \frac{d^2\sigma}{d\xbj\, d\ybj } =
  \frac{4\pi\alpha^2}{S\xbj^2\ybj^2} \left[ \left\{ 1-\ybj +\frac{\ybj^2}{2}
%-\frac{M^2 \xbj \ybj }{S-M^2}{\rm parton\ k_{1}}(k_{1})
    \right\} F_2 (\xbj ,Q^2) - \frac{\ybj^2}{2} F_L (\xbj ,Q^2 ) \right] ,
\end{equation}
are expressed via
a standard convolution in terms of the parton structure functions 
${\mathcal F}_{i\,{p}}$ and corresponding parton densities $f_{ p}$,
\begin{equation}
  F_i\, (\xbj ,Q^2) = \sum_{p=q,g} \int_{\xbj}^1 \frac{dx}{x}\, f_{ p}\left(
    \frac{\xbj}{x}\right) \, \frac{\xbj}{x}\, 
 {\mathcal F}_{i\,{ p}} (x,Q^2) 
\ ,\ 
  \ i=2,L\, .
\label{protonstruc}
\end{equation}  
Here, $\sqrt{S}$ is the center-of-mass (c.m.) energy of the 
lepton-hadron
system. The corresponding Bjorken variables are defined as usual
\begin{equation}
  \xbj \equiv \frac{Q^2}{2\,P\cdot q} ; \hspace{48pt} \ybj \equiv \frac{P\cdot 
    q}{P\cdot k} ,
\end{equation}
where $P\,(k)$ is the four-momentum of the incoming proton (lepton).

%%%%%%%%%%%%%%%%%%%%%%%%%%%%%FIGURE  %%%%%%%%%%%%%%%%%%%%%%%%%%%%%%%%%%%
\begin{figure} 
\begin{center}
  \epsfig{file=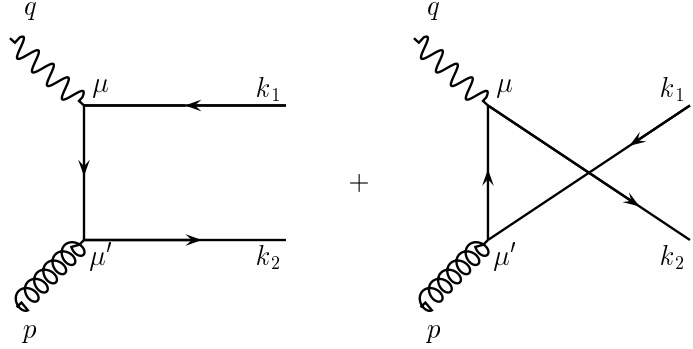,width=9cm}
\caption[dum]{\label{f2}
  Perturbative {\it chirality-conserving} process, $\gamma^{\ast}(q)+{\rm
    g}(p )\rightarrow \overline{{\rm q}_{L}}(k_{1})+{\rm
    q}_{L}(k_{2})$.  }
\end{center}
\end{figure}
%%%%%%%%%%%%%%%%%%%%%%%%%%%%%%%%%%%%%%%%%%%%%%%%%%%%%%%%%%%%%%%%%%%%%% 

As outlined in the Introduction, we shall consider in this paper only 
the contributions from the simplest
instanton-induced, {\it chirality-violating} photon-gluon process, 
corresponding to one massless quark flavour $(n_{f}=1)$,
\begin{equation}
  \gamma^{\ast}(q)+{\rm g}(p)\rightarrow 
   \overline{{\rm q}_{L}}(k_{1})+{\rm  q}_{R}(k_{2})\ ;
\hspace{24pt}
\left( \triangle
Q_{5}\equiv \triangle \left( Q_{R}-Q_{L}\right)=2\right) \, .
\label{instproc}
\end{equation}
It will be very instructive to compare with the appropriate leading-order
perturbative QCD amplitudes for the {\it chirality-conserving} process
(Fig.~\ref{f2}),
\begin{equation}
  \gamma^{\ast}(q)+{\rm g}(p)\rightarrow 
     \overline{{\rm q}_{L}}(k_{1})+{\rm q}_{L}(k_{2})\ ;
\hspace{24pt}
\left( \triangle
Q_{5}\equiv \triangle \left( Q_{R}-Q_{L}\right)=0\right) \, .
\label{pertproc}
\end{equation} 
at the various stages of the instanton calculation. Therefore, for reference,
let us summarize the well-known perturbative results next 
(see any textbook on perturbative QCD, e.g. Ref.~\cite{f}). 

Let us use two-component Weyl-notation for the (massless) fermions involved, 
in order to facilitate the comparison with the instanton calculation later on.
The leading-order amplitude for the perturbative process (\ref{pertproc})
(Fig.~\ref{f2}) then reads,
\begin{eqnarray}
\label{ampl_pt}
\lefteqn{{\mathcal T}^{A}_{\mu \,\mu ^{\prime}}
     \left( \gamma^{\ast}+{\rm g}\rightarrow
    \overline{{\rm q}_{L}}+{\rm q}_{L}\right) =} 
\\[0.5ex]
\nonumber
&& 
e_{q}\,g_s\,t^{A}\,  \chi_{L}^{\dagger}(k_{2}) \left[
    \overline{\sigma}_{\mu ^{\prime}}\frac{(q-k_{1})}{(q-k_{1})^{2}}
     \overline{\sigma}_\mu 
       -\overline{\sigma}_\mu \frac{(q-k_{2})}{(q-k_{2})^{2}}
      \overline{\sigma}_{\mu ^{\prime}} \right]
  \chi_{L}(k_{1})\ ,
\end{eqnarray}
where the two-component Weyl-spinors $\chi_{L,\,R}$ satisfy
the Weyl-equations, 
\begin{equation}
\overline{k}\,\chi_L(k)=0 \, ;\hspace{24pt}
k\,\chi_R(k) =0 \, ,
\label{weyleq}
\end{equation}
and 
\begin{equation}
  \chi_L(k)\,\chi_L^\dagger (k) = k \, ; \hspace{24pt}
  \chi_R(k)\,\chi_R^\dagger (k) = \overline{k} \, .
\label{compl}   
\end{equation}
In Eqs.~(\ref{ampl_pt}-\ref{compl}) and throughout the paper we use the
abbreviations,
\begin{equation}
  v \equiv v_\mu \,\sigma^\mu \, ; \hspace{24pt} \overline{v} \equiv v_\mu 
  \,\overline{\sigma}^\mu \, ,\ \mbox{for any four-vector $v_\mu $},
\end{equation}
where the familiar $\sigma$-matrices\footnote{\label{sigmas} 
We use the standard notations, in Minkowski space: 
$\sigma_\mu =(1,\vec{\sigma})$, $\overline{\sigma}_\mu =(1,-\vec{\sigma})$, 
and in Euclidean space: 
$\sigma_\mu =(-\ii\,\vec{\sigma},1)$, $\overline{\sigma}_\mu =
(\ii\,\vec{\sigma},1)$, where $\vec{\sigma}$ are the Pauli
matrices.}
 satisfy,
\begin{equation}
  \sigma_\mu \overline{\sigma}_\nu +\sigma_\nu \overline{\sigma}_\mu =
  2\,g_{\mu \nu} \, .
\label{com}
\end{equation}
Finally, in Eq.~(\ref{ampl_pt}), $t^A,A=1,\ldots ,8$, are the SU(3)
generators, $e_q$ is the quark charge in units of the electric charge $e$, and
$g_s$ is the SU(3) gauge coupling.

With help of Eqs.~(\ref{weyleq}), (\ref{com}) and the on-shell conditions 
$k_1^2=k_2^2=0$, the gauge-invariance constraints,
\begin{equation}
  q^\mu \,{\mathcal T}^{A}_{\mu \,\mu ^{\prime}}=0\, ;\hspace{24pt} {\mathcal
    T}^{A}_{\mu \,\mu ^{\prime}}\,p^{ \mu ^{\prime}}=0\, ,
\label{gaugeinv}
\end{equation}
are easily checked. 

Next, we obtain the leading-order contribution 
of the process (\ref{pertproc}) to ${\mathcal W}^{\mu \nu\,(2)}_{g}(q,p)$ by 
contracting Eq.~(\ref{ampl_pt}) with the gluon
polarization vector $\epsilon_{g}^{\mu ^{\prime}}(p)$ and 
taking the traces in Eq.~(\ref{wmunu_n}) by means of
relations (\ref{compl}) and (\ref{com}). Averaging over the initial-state 
gluon polarization and colour amounts to an overall factor $1/16$.
The final result for the projections needed in 
Eqs.~(\ref{diffcross}), (\ref{projF2}), and (\ref{projFL}) 
then reads
\begin{eqnarray}
\label{gmunupt}
\lefteqn{ -g_{\mu \nu}\, 
\frac{d{\mathcal W}^{\mu \nu\,{(2)}}_{g}}{dt}
\,\left( \gamma^{\ast}+{\rm g}\rightarrow
    \overline{{\rm q}_{L}}+{\rm q}_{L}\right) 
=}
\\[0.5ex]
\nonumber
&&  
e_q^2\, \frac{\alpha_s}{4\,\pi}\,
\frac{x}{2\,Q^2}\, \left[
    \frac{u}{t}+\frac{t}{u}-2\,\frac{1-x}{x}\,\frac{Q^4}{tu}
  \right]
\, ,
\end{eqnarray}
\begin{equation}
\label{pmupnupt}
 \frac{p_\mu \,p_{\nu}}{p\cdot q} \, 
\frac{d{\mathcal W}^{\mu \nu\,{(2)}}_{g}}{dt}
\,\left( \gamma^{\ast}+{\rm g}\rightarrow
    \overline{{\rm q}_{L}}+{\rm q}_{L}\right)
=
e_q^2\,\frac{\alpha_s}{4\,\pi}\,
\frac{x\,(1-x)}{Q^2}
\, ,          
\end{equation}
where $u=-t-Q^2/x$.

Upon integrating Eq.~(\ref{gmunupt}) over $t$, we encounter the familiar 
collinear divergencies for $t,u\rightarrow 0$. In order to isolate the hard 
contributions to the gluon structure functions, it is adequate to regularize 
the collinear singularities by introducing an infrared cutoff scale 
$\mu _{c}$ in the integration limits,   
$\{-Q^{2}/x+\mu _{c}^{2},-\mu _{c}^{2}\}$. On account of 
Eqs.~(\ref{projF2}), (\ref{projFL}), one then 
obtains the familiar results\footnote{\label{foot1}When comparing with the
literature, one has to remember that we considered  only the production of a 
$\overline{{\rm q}_{L}}\,{\rm q}_{L}$ pair (c.\,f. Eq.~(\ref{pertproc})). 
In the full ${\mathcal O}(\alpha_{s})$ contribution
to the gluon structure functions, the production of a
$\overline{{\rm q}_{R}}\,{\rm q}_{R}$ pair has also to be included. This
amounts to  multiplying Eqs.~(\ref{F2pert}), 
(\ref{FLpert}), by a
factor of 2.} in the Bjorken limit, 
\begin{eqnarray}
\label{F2pert} 
&&  {\mathcal F}_{2\,g}^{(\overline{{\rm q}_L}{\rm q}_L)}
         \,  (x,Q^2;\mu _{c}^{2})=
e_q^2\, \frac{\alpha_s}{2\,\pi}\,\times 
\\[0.8ex]
\nonumber
&& 
x\,
   \left[
          P_{qg}(x)\,
    \ln\left(\frac{Q^{2}}{\mu _{c}^{2}}\right)
          +P_{qg}(x)\,\ln \left(\frac{1}{x}\right) 
-\frac{1}{2} +3\,x\,(1-x)
  \right]
\left[1+{\mathcal O}\left(\frac{\mu _{c}^{2}}{Q^{2}}\right)\right]
 \, ,
\end{eqnarray}
\begin{equation}
{\mathcal F}_{L\,g}^{(\overline{{\rm q}_L}{\rm q}_L)}\,(x,Q^2)=
e_q^2\,\frac{\alpha_s}{\pi}\,
x^{2}\,(1-x)
\, ,
\label{FLpert}
\end{equation}
with the splitting function
\begin{equation}
\label{pqg}
P_{qg}(x) \equiv \frac{1}{2}\,\left(x^{2}+(1-x)^{2}\right)\, .
\end{equation}
Of course, the {\it finite} part of the structure function 
${\mathcal F}_{2\,g}^{(\overline{{\rm q}_L}{\rm q}_L)}$, 
that is everything except for the large logarithm, 
$\ln (Q^{2}/\mu _{c}^{2})$, is scheme dependent.

\section{The Instanton-Induced Process\\ $\gamma^\ast + {\rm g}\rightarrow
  \overline{{\rm q}_L} + {\rm q}_R$}

In this section, we turn to the central issue of this paper. We  
consider the simplest instanton-induced 
exclusive process, $\gamma^\ast + {\rm g}\rightarrow\overline{{\rm q}_L} +
{\rm q}_R$,  and compute its  contributions to the fixed angle differential 
cross-section and the gluon structure functions  
${\mathcal F}_{2\,{ g}}$ and 
${\mathcal F}_{L\,{ g}}$, in leading semi-classical approximation.

To this end, the respective Green's function is first
set up according to  standard instanton-perturbation theory in Euclidean 
configuration space~\cite{th,brown,ber,abc,r}, then Fourier transformed to 
momentum space, LSZ amputated, and finally continued to Minkowski space. 

The basic building blocks are (in Euclidean 
configuration space and in the singular gauge):  

\bigskip
\noindent
i) The classical instanton gauge field~\cite{bpst} 
$A^{(I)}_{\mu ^{\prime}}$,
\begin{eqnarray}
A_{\mu ^{\prime}}^{(I)}(x)&=&-\ii\,\frac{2\,\pi^{2}}{g_{s}}\,
\rho^{2}\,U\,\left(
\frac{
\sigma_{\mu ^{\prime}}\,\overline{x}-x_{\mu ^{\prime}}}{2\,\pi^{2}\,x^{4}}
\right)\,U^{\dagger}
\ 
\frac{1}{\Pi_{x}}\, ,
\label{igauge}
\\
\Pi_x &\equiv & 1+\frac{\rho^2}{x^2} \, ,
\end{eqnarray}
depending on the various collective coordinates, the instanton size $\rho$
and the colour orientation matrices $U^k_{\ \alpha}$. The $U$ matrices  
involve both  colour ($k=1,2$) and spinor ($\alpha=1,2$) indices, the
former ranging  as usual only in the 
$2\times 2$ upper left corner of $3\times 3$ SU(3) colour matrices. 
Indices will, however, be suppressed, as long as no confusion can arise.
  
\bigskip
\noindent
ii) The quark zero modes~\cite{th}, $\kappa$ and $\overline{\phi}$,
\begin{eqnarray}
\kappa^m_{\ \dot\beta}\,(x)
&=& 2\,\pi\,\rho^{3/2}\, \epsilon^{\gamma\delta}\,
\left( U\right)^m_{\ \delta}\,
\frac{\overline{x}_{\dot\beta\gamma}}{2\,\pi^2\,x^4}\ 
\frac{1}{\Pi_x^{3/2}}\, ,
\label{kappa}
\\
\overline{\phi}^{\dot\alpha}_{\ l}\,(x)
&=&2\,\pi\,\rho^{3/2}\, \epsilon_{\gamma\delta}\,
\left( U^\dagger\right)^\gamma_{\ l}\,
\frac{x^{\delta\dot\alpha}}{2\,\pi^2\,x^4}\ 
\frac{1}{\Pi_x^{3/2}}\, ,
\label{phibar}
\end{eqnarray}
and

\bigskip
\noindent
iii) 
the quark propagators in the instanton background~\cite{brown},
\begin{eqnarray}
\label{si}
&&S^{(I)}(x,y) =
\\[1.6ex]
\nonumber
&&
\frac{1}{\sqrt{\Pi_x\Pi_y}}
\left[ 
\frac{x-y}{2\pi^2(x-y)^4}
\left( 1 +\rho^2\frac{
\left[ U x \overline{y} U^\dagger\right]}
{x^2 y^2}\right)
+\frac{\rho^2\sigma_\mu }{4\pi^2}
\frac{\left[
U x\,(\overline{x}-\overline{y}) \sigma_\mu \overline{y} U^\dagger
\right]}
{x^2(x-y)^2 y^4\Pi_y}
\right],
\\[2ex]
\label{sibar}
&&\overline{S}^{(I)}(x,y) =
\\[1.6ex]
\nonumber
&&
\frac{1}{\sqrt{\Pi_x\Pi_y}}
\left[ 
\frac{\overline{x}-\overline{y}}{2\pi^2(x-y)^4}
\left(  1 +\rho^2\frac{
\left[ U x \overline{y} U^\dagger\right]}
{x^2 y^2}\right)
+\frac{\rho^2 
\overline{\sigma}_\mu }{4\pi^2}
\frac{\left[
U x \overline{\sigma}_\mu (x-y) \overline{y} 
U^\dagger \right]}
{\Pi_x x^4(x-y)^2 y^2  }
\right] .
\end{eqnarray}

%%%%%%%%%%%%%%%%%%%%%%%%%%%%%FIGURE  %%%%%%%%%%%%%%%%%%%%%%%%%%%%%%%%%%%
\begin{figure}
\begin{center}
  \epsfig{file=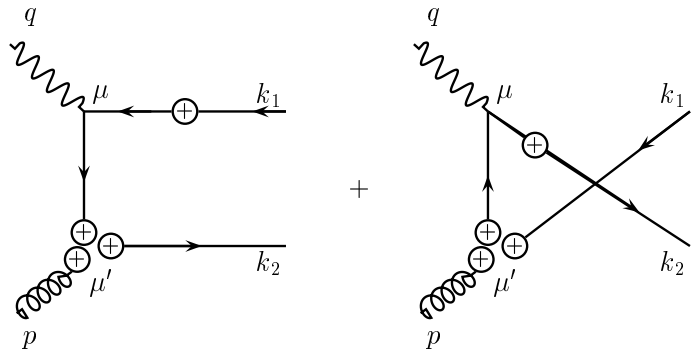,width=9cm}
\caption[dum]{\label{f3}
  Instanton-induced {\it chirality-violating} process, $\gamma^{\ast}(q)
  +{\rm g}(p )\rightarrow \overline{{\rm q}_{L}}(k_{1})+{\rm
    q}_{R}(k_{2})$, in leading semi-classical approximation. The corresponding
  Green's function involves the products of the appropriate classical fields
  (lines ending at blobs) as well as the quark propagator in the instanton
  background (quark line with central blob).  }
\end{center}
\end{figure}
%%%%%%%%%%%%%%%%%%%%%%%%%%%%%%%%%%%%%%%%%%%%%%%%%%%%%%%%%%%%%%%%%%%%%% 
     
The relevant diagrams for the exclusive process of interest,
Eq.~(\ref{instproc}), are displayed in Fig.~\ref{f3}, in leading semi-classical
approximation. The amplitude is expressed in terms of an
integral over the collective coordinates $\rho$ and
the colour orientation $U$,
\begin{equation}
  {\mathcal T}_{\mu \,\mu ^{\prime}}^{a} 
     \left( \gamma^\ast + {\rm g}\rightarrow
    \overline{{\rm q}_L} + {\rm q}_R \right) = \int\limits_0^\infty
  \frac{d\rho}{\rho^5}\,d(\rho ,\mu _{r} )\,\int dU\, {\mathcal A}_{\mu 
    \,\mu ^{\prime}}^{a}(\rho, U) \,;\ a=1,2,3,
\label{ampl_inst}
\end{equation}
where
\begin{equation}
  d(\rho ,\mu _{r} )= d\,\left( \frac{2\,\pi}{\alpha_s(\mu _{r} )}\right)^6\, 
     {\exp}\left[-\frac{2\,\pi}{\alpha_s(\mu _{r} )}\right] \,
     \left(
    \rho\,\mu _{r} \right)^{\beta_0+\frac{\alpha_s(\mu _{r} )}{4\,\pi}\, 
     \left( \beta_1 - 12\,\beta_0 \right)}\, ,
\label{density}
\end{equation}
denotes the instanton density~\cite{th,ber,abc,morretal}, with $\mu _{r}$ 
being the
renormalization scale. The form (\ref{density}) of the density, with
next-to-leading order expression for $\alpha_{s}(\mu _{r})$, 
\begin{equation}
\label{alpha}
\alpha_{s}(\mu _{r})=
\frac{4\,\pi}{\beta_{0}\,\ln \left(\frac{\mu _{r}^{2}}{\Lambda^{2}}\right)}
\left[ 1 - \frac{\beta_{1}}{\beta_{0}^{2}}
\frac{\ln\left(\ln \left(\frac{\mu _{r}^{2}}{\Lambda^{2}}\right) \right)}
{\ln \left(\frac{\mu _{r}^{2}}{\Lambda^{2}}\right)}
\right] \, ,
\end{equation}
is improved to
satisfy renormalization-group invariance at the 2-loop
le\-vel~\cite{morretal}.  The constants $\beta_{0}$ and $\beta_{1}$ 
are the familiar perturbative coefficients of the QCD beta-function,
\begin{equation}
\label{beta}
  \beta_0=11-\frac{2}{3}\,n_f\, ;\hspace{24pt} 
  \beta_1=102-\frac{38}{3}\,n_f \, ,
\end{equation}
and the constant $d$ is given by\footnote{Strictly speaking, the constant
$d$ is known only to $1$-loop accuracy. In Ref.~\cite{morretal}, only
the ultraviolet divergent part of the $2$-loop correction to the 
instanton density has been computed.}
\begin{equation}
\label{d}
  d=\frac{C_1}{2}\,{\rm e}^{-3\,C_2+n_f\,C_3}\, ,
\end{equation}
with $C_{1}=0.466$, $C_{2}=1.54$, and $C_{3}=0.153$, in the $\overline{\rm
  MS}$-scheme. 
In our case, we should of course take $n_{f}=1$ in Eqs.~(\ref{beta}) and 
(\ref{d}).

Before analytic continuation, the amplitude 
${\mathcal A}_{\mu \,\mu ^{\prime}}$
entering Eq.~(\ref{ampl_inst}) takes the following form in Euclidean space,
\vfill\eject
\begin{eqnarray}
  \label{ampi}
\lefteqn{  {\mathcal A}^{a}_{\mu \,\mu ^{\prime}} = -\ii\,e_{q}\,
    \lim_{p^{2}\to 0}p^{2}\,
    {\rm tr}\left( \sigma^{a}\,A_{\mu ^{\prime}}^{(I)}(p)\right)\times }  
\\ 
\nonumber
&&
\chi_{R}^{\dagger}(k_{2})\, 
   \left[ 
     \lim_{k_{2}^{2}\to 0}\,(\ii k_{2})\,\kappa (-k_{2})\,
     {{\mathcal V}_\mu ^{(t)}}(q,-k_{1})\right .
\\
\mbox{}&&+  
\left .    
{{\mathcal V}_\mu ^{(u)}}(q,-k_{2})\,\lim_{k_{1}^{2}\to 0}\,
    \overline{\phi}(-k_{1})\,(-\ii\,\overline{k}_{1}) 
\right] 
\chi_{L}(k_{1})\, ,
\nonumber
\end{eqnarray}
with contributions ${{\mathcal V}_\mu ^{(t,u)}}$ from the diagrams 
on the left and right in Fig.~\ref{f3}, respectively, 
\begin{eqnarray}
{{\mathcal V}_\mu ^{(t)}}(q,-k_{1})&\equiv&
\int d^{4}x\,{\rm e}^{-\ii\,q\cdot x} \,
\left[ 
\overline{\phi}(x)\,\overline{\sigma}_\mu \,
\lim_{k_{1}^{2}\to 0}\,S^{(I)}\,(x,-k_{1})\,(-\ii\,\overline{k}_{1})\,
\right] ,
\label{tinst}\\
{{\mathcal V}_\mu ^{(u)}}(q,-k_{2})&\equiv&
\int d^{4}x\,{\rm e}^{-\ii\,q\cdot x} \,
\left[
\lim_{k_{2}^{2}\to 0}\,(\ii k_{2})\,\overline{S}^{(I)}\,(-k_{2},x)\,
\sigma_\mu \,\kappa(x)\right]\, ,
\label{uinst}
\end{eqnarray}
and generic notation for various Fourier transforms involved, 
\begin{equation}
\label{fourier}
f(\ldots,k,\ldots)=\int d^4 x\,{\rm e}^{-\ii\,k\cdot x}\,
f(\ldots,x,\ldots)\, .
\end{equation}

The LSZ-amputation of the classical instanton 
gauge field $A^{(I)}_{\mu ^{\prime}}$
in Eq.~(\ref{ampi}) and the quark zero modes $\kappa$ and $\overline{\phi}$
in Eqs.~(\ref{tinst}) and (\ref{uinst}), respectively, 
is straightforward~\cite{r},  
\begin{eqnarray}
\lim_{p^{2}\to 0}p^{2}\,
{\rm tr}\left( \sigma^{a}\,A_{\mu ^{\prime}}^{(I)}(p)\right) &=&
\frac{2\,\pi^{2}}{g_{s}}\,\rho^{2}\,
{\rm tr}\left[ \sigma^{a}\,U\,\left[
p_{\mu ^{\prime}}-\sigma_{\mu ^{\prime}}\,\overline{p}\right]\,U^{\dagger}
\right]\, ,
\label{lszgluon}\\
\lim_{k_{2}^{2}\to 0}\,
(\ii k_{2})^{\alpha\dot\alpha}\,\kappa^{i}_{\dot\alpha}(-k_{2})&=&
2\,\pi\,\rho^{3/2}\,U^{i}_{\ \beta}\,\epsilon^{\beta\alpha}\, ,
\label{lszkappa}\\
\lim_{k_{1}^{2}\to 0}\,\overline{\phi}_{j}^{\dot\gamma}(-k_{1})\,
(-\ii\,\overline{k}_{1})_{\dot\gamma\delta}
&=&2\,\pi\,\rho^{3/2}\,\epsilon_{\beta\delta}\,
\left( U^{\dagger}\right)^{\beta}_{\ j}\, .
\label{lszbarphi} 
\end{eqnarray}

On the other hand, the LSZ-amputation of the quark propagators 
$S^{(I)}$ and $\overline{S}^{(I)}$ in 
Eqs.~(\ref{tinst}) and (\ref{uinst}), respectively, 
is quite non-trivial and has important physical consequences, as we
shall see below. We give here only the final result and refer the interested
reader to Appendix A where the details of the calculation can be 
found:
\vfill\eject
\begin{eqnarray}
\label{onshellsi}
\lefteqn{ \lim_{k_1^2\to 0}\,
S^{(I)}\,(x,-k_1)\,(-\ii\,
\overline{k_1}) =}
\\[0.8ex]
\nonumber
&& 
\frac{- 1}{\sqrt{\Pi_x}}\,
{\rm e}^{\ii\,k_1\cdot x}\,
\left[
 1 +
\frac{1}{2}\,\frac{\rho^2}{x^2}\,
\frac{\left[ U\,x\,\overline{k_1}\,U^\dagger
\right]}{k_1\cdot x}\,
\left( 1- {\rm e}^{-\ii\,k_1\cdot x}\right)
\right]
\, ,
\\[1.6ex]
\label{onshellsibar}
\lefteqn{
\lim_{k_2^2\to 0}\,
\left( \ii\,k_2\right)\,
\overline{S}^{(I)}\,(-k_2,x)
=}
\\[0.8ex]
\nonumber
&&
\frac{- 1}{\sqrt{\Pi_x}}\,
{\rm e}^{\ii\,k_2\cdot x}\,
\left[
 1 +
\frac{1}{2}\,\frac{\rho^2}{x^2}\,
\frac{\left[ U\,k_2\,\overline{x}\,U^\dagger
\right]}{k_2\cdot x}\,
\left( 1- {\rm e}^{-\ii\,k_2\cdot x}\right)
\right]
\, .
\end{eqnarray}
It should be noted that the first terms in 
Eqs.~(\ref{onshellsi}), (\ref{onshellsibar}), corresponding to the 
$1$ in square brackets, were argued to be present on general grounds 
already in Ref.~\cite{bbb}. The remaining terms, however, have
not been given in the literature. As we shall see below, 
they play a very important r{\^o}le in ensuring electromagnetic
gauge invariance. 

The Fourier transforms entering Eqs.~(\ref{tinst}) and (\ref{uinst}),
respectively,
can now be done with the help of Eqs.~(\ref{onshellsi}) and 
(\ref{onshellsibar}). The result is (see Appendix B),
\begin{eqnarray}
\label{tvertex}
\lefteqn{
{\mathcal V}^{(t)}_{\mu \ j\,\alpha}\,(q,-k_{1}) 
=
2\,\pi\,\ii\,\rho^{3/2}\,
\left( U^{\dagger}\right)^\gamma_{\ j}\, 
 \left\{
\frac{1}{2}\,
\frac{\left[ \epsilon\,k_{1}\,\overline{\sigma}_\mu \right]_{\gamma\alpha}}
{q\cdot k_{1}}\, f\left( \rho\,\sqrt{q^{2}} \right) \right .}
\\[1.6ex]
\nonumber
&&
\left .
+ \left[
\frac{\left[ \epsilon\,\left(q-k_{1}\right) 
\,\overline{\sigma}_\mu \right]_{\gamma\alpha}}
{(q-k_{1})^2}
-\frac{1}{2}\,
\frac{\left[ \epsilon\,k_{1}\,\overline{\sigma}_\mu \right]_{\gamma\alpha}}
{q\cdot k_{1}}
\right]\,f\left(\rho\,\sqrt{\left( q-k_{1}\right)^2}\right)
\right\}\, ,
\nonumber 
\\[2.4ex]
\label{uvertex} 
\lefteqn{
{\mathcal V}^{(u)\ \alpha\, i}_{\mu \ }\,(q,-k_{2}) =
2\,\pi\,\ii\,\rho^{3/2}\,
U^i_{\ \gamma}\,
 \left\{
\frac{1}{2}\,
\frac{\left[ \sigma_\mu \,
\overline{k_{2}}\,\epsilon\right]^{\alpha\gamma}}
{q\cdot k_{2}}\, f\left( \rho\,\sqrt{q^{2}}\right) \right .}
\\[1.6ex]
\nonumber
&&
\left .
+ \left[
\frac{\left[ \sigma_\mu \,
\left(\overline{q}-\overline{k_{2}}\right)\,\epsilon 
\right]^{\alpha\gamma}}
{(q-k_{2})^2}
-\frac{1}{2}\,
\frac{\left[ \sigma_\mu \,
\overline{k_{2}}\,\epsilon\right]^{\alpha\gamma}}
{q\cdot k_{2}}
\right]\,f\left(\rho\,\sqrt{\left( q-k_{2}\right)^2}\right)
\right\}\, ,
\nonumber 
\end{eqnarray}
with the shorthand (``form factor''), 
\begin{equation}
\label{form}
f(\omega )\equiv \omega\,K_{1}(\omega ), 
\end{equation}
in terms of the Bessel-K function, implying the normalization,
\begin{equation}
\label{formnorm}
f(0)=1\, .
\end{equation}

The next step is to insert Eqs.~(\ref{tvertex}), (\ref{uvertex}), 
and (\ref{lszgluon}-\ref{lszbarphi}) into Eq.~(\ref{ampi}) and to
perform the integration over the colour orientation according to
Eq.~(\ref{ampl_inst}) by means of the relation,
\begin{eqnarray}
\label{colorint}
\lefteqn{
\int dU\,U^k_{\ \beta^\prime}\,(U^\dagger )^{\gamma^\prime}_{\ l}\,
U^i_{\ \tau}\,(U^\dagger )^\gamma_{\ j} =}
\\ 
\nonumber 
&&\frac{1}{6}\,\left[ 
\delta_\tau^{\ \gamma^\prime}\,\delta^i_{\ l}\ 
\delta_{\beta^\prime}^{\ \gamma}\,\delta^k_{\ j}
+
\delta_\tau^{\ \gamma}\,\delta^i_{\ j}\ 
\delta_{\beta^\prime}^{\ \gamma^\prime}\,\delta^k_{\ l}
+
\epsilon_{\tau\,\beta^\prime}\,\epsilon^{i\,k}\
\epsilon^{\gamma^\prime\,\gamma}\,\epsilon_{l\,j} 
\right] \, .
\end{eqnarray}
After analytic continuation to Minkowski space we find for the
scattering amplitude, Eq.~(\ref{ampl_inst}),
\begin{eqnarray}
\label{ampi2}
&& {\mathcal T}_{\mu \,\mu ^{\prime}}^a\,
   \left( \gamma^\ast + {\rm g}\rightarrow
    \overline{{\rm q}_L} + {\rm q}_R \right)   
 = -\ii\,\frac{4}{3}\,\pi^4\,\frac{e_q}{g_s}\,\sigma^a \int\limits_0^\infty
  d\rho\,d(\rho ,\mu _{r} )\,\times 
\\[0.8ex]
&&
\chi_R^\dagger (k_2)
\left[ \left(
\sigma_{\mu ^{\prime}}\overline{p}-p\overline{\sigma}_{\mu ^\prime}\right)
V(q,k_1;\rho )\overline{\sigma}_\mu 
-\sigma_\mu \overline{V}(q,k_2;\rho )
\left(
\sigma_{\mu ^\prime}\overline{p} - p\overline{\sigma}_{\mu ^\prime}
\right)
\right]
\chi_L(k_1) ,
\nonumber
\end{eqnarray}
with the four-vector $V_{\lambda}$,
\begin{eqnarray}
\nonumber
  V_\lambda (q,k;\rho ) &\equiv& 
  \left[  
\frac{\left( q-k\right)_\lambda}{-(q-k)^2}
+\frac{k_{\lambda}}{2 q\cdot k}
\right]\rho\sqrt{-\left( q-k\right)^2}\,
K_1\left(\rho\sqrt{-\left( q-k\right)^2}\right)
\\[0.8ex]
\label{V}
\mbox{}&&
-
\frac{k_{\lambda}}{2 q\cdot k}\rho\sqrt{-q^{2}}\,
K_1\left(\rho\sqrt{-q^{2}}\right) . 
\end{eqnarray}

At this stage of our instanton calculation, 
the gauge-invariance con\-st\-raints, Eqs.~(\ref{gaugeinv}),
can easily be checked. While the relation 
${\mathcal T}_{\mu \,\mu ^{\prime}}^a\,p^{\mu ^\prime}=0$ holds trivially, 
the electromagnetic (e.m.) current conservation 
$q^\mu \,{\mathcal T}_{\mu \,\mu ^{\prime}}^a=0$ follows again 
from the relations (\ref{com}) of the $\sigma$-matrices, the Weyl-equations 
(\ref{weyleq}) and the on-shell conditions $k_1^2=k_2^2=0$. 
Electromagnetic current conservation provides also for a non-trivial
check of our result for the amputated quark propagators, which differs
somewhat from the result quoted in Ref.~\cite{bbb}: 
If we keep only the first terms in 
Eqs.~(\ref{onshellsi}), (\ref{onshellsibar}), corresponding to the 
$1$ in square brackets, e.m. current conservation would only hold
for a restricted set of momenta in phase space, namely for 
$(q-k_1)^2=(q-k_2)^2$.       

Furthermore, we note one of the main results of this paper:
The integration over the instanton size $\rho$ in Eq.~(\ref{ampi2})
is {\it finite}. In particular, the {\it good infrared} behavior 
(large $\rho$) of the integrand is due to the exponential decrease 
of the Bessel-K function for large $\rho$ in Eq.~(\ref{V}). Its origin, 
in turn, can be traced back to the ``feed-through'' of the factor
$1/\sqrt{\Pi_x}$, by which the amputated (current) quark propagators
(\ref{onshellsi}) and (\ref{onshellsibar}) in the $I$-background 
essentially differ from the respective amputated 
free propagators. If the current-quark propagators 
in Eqs.~(\ref{tinst}) and (\ref{uinst}) are naively approximated by the 
free ones (c.\,f. Eqs.~(\ref{si}), (\ref{sibar})), 
\begin{equation}
S^{(0)}(x,y)=\frac{x-y}{2\pi^2(x-y)^4}\,;\hspace{24pt}
\overline{S}^{(0)}(x,y)=\frac{\overline{x}-\overline{y}}{2\pi^2(x-y)^4}\, ,
\end{equation}
the result is both gauge variant and contains an IR-divergent 
piece in the $\rho$ integration. 

We have thus demonstrated explicitly and to our
knowledge for the first time that the typical hard scales ($Q^{2},\ldots$) 
in deep-inelastic scattering provide a dynamical IR cutoff for the
instanton size (at least in leading semi-classical approximation).

Now we are ready to perform the final integration over the instanton
size $\rho$ by inserting the instanton density, Eq.~(\ref{density}), into
Eq.~(\ref{ampi2}). The result is: 
\begin{eqnarray}
\label{ampifin}
&& {\mathcal T}_{\mu \,\mu ^{\prime}}^a\,
  \left( \gamma^\ast + {\rm g}\rightarrow
    \overline{{\rm q}_L} + {\rm q}_R \right)   
   = 
\\[2.4ex]
\nonumber
&& -\ii\frac{\sqrt{2}}{3}d\pi^{3}e_q
     \left( \frac{2\,\pi}{\alpha_s(\mu _{r} )}\right)^{13/2}
    {\exp}\left[-\frac{2\,\pi}{\alpha_s(\mu _{r} )}\right]\,
   2^{\,b}\Gamma \left(\frac{b+1}{2}\right)
   \Gamma \left(\frac{b+3}{2}\right)
\sigma^a\,\times 
\\[2.4ex]
&&
\chi_R^\dagger (k_2)
\left[ \left(
\sigma_{\mu ^{\prime}}\overline{p}-p\overline{\sigma}_{\mu ^\prime}\right)
v(q,k_1;\mu _{r} )\overline{\sigma}_\mu 
-\sigma_\mu \overline{v}(q,k_2;\mu _{r} )
\left(
\sigma_{\mu ^\prime}\overline{p} - p\overline{\sigma}_{\mu ^\prime}
\right)
\right]
\chi_L(k_1) ,
\nonumber
\end{eqnarray}
with the four-vector $v_{\lambda}$,
\begin{eqnarray}
  \label{v}  
&&v_\lambda (q,k;\mu _{r} ) \equiv 
\\[1.6ex]
\nonumber
&& 
\frac{1}{\mu _{r}}\,\left\{
  \left[  
\frac{\left( q-k\right)_\lambda}{-(q-k)^2}
+\frac{k_{\lambda}}{2 q\cdot k}
\right] \left(
\frac{\mu _{r}}{\sqrt{-\left( q-k\right)^2}}\right)^{b+1}
-
\frac{k_{\lambda}}{2 q\cdot k}
\left( \frac{\mu _{r}}{\sqrt{-q^{2}}}\right)^{b+1} 
\right\} . 
\end{eqnarray}
In Eqs.~(\ref{ampifin}) and  (\ref{v}), the variable $b$ is a shorthand for the
effective power of $\rho\mu _{r}$ in the instanton density, 
Eq.~(\ref{density}),
\begin{equation}
\label{beff}
b\equiv
\beta_0+\frac{\alpha_s(\mu _{r} )}{4\,\pi}\, \left( \beta_1 -
      12\,\beta_0 \right) \, .
\end{equation}

The next steps consist in contracting the amplitude, Eq.~(\ref{ampifin}),
with the gluon polarization vector $\epsilon_{g}^{\mu ^{\prime}}(p)$ and  
taking the modulus squared of this amplitude according to Eq.~(\ref{wmunu_n}).
After applying Eq.~(\ref{compl}), 
the remaining spinor traces can be evaluated, in 
principle, by repeated use of Eq.~(\ref{com}). For the actual calculation,
however, we used FORM and, for an independent check, the HIP package 
for MAPLE. The final result for the relevant projections  
(c.f. Eqs.~(\ref{diffcross}), (\ref{projF2}), and (\ref{projFL})) 
of the contribution of the $I$-induced process (\ref{instproc}) 
to the differential gluon structure tensor is found to be
\begin{eqnarray}
\nonumber
\lefteqn{
 -g_{\mu \nu}\, \frac{d{\mathcal W}^{\mu \nu\,{(2)}}_{g}}{dt}
\, \left( \gamma^\ast + {\rm g}\rightarrow
    \overline{{\rm q}_L} + {\rm q}_R \right)   
 =}
\\[1.6ex]
\label{gmunu}
&&\frac{e_q^2}{16}\,
   {\mathcal N}^{2}\,
\left( \frac{2\,\pi}{\alpha_s(\mu _{r} )}\right)^{13}\,
{\exp}\left[-\frac{4\,\pi}{\alpha_s(\mu _{r} )}\right]\,
\left( \frac{\mu _{r}^{2}}{Q^{2}}\right)^{b}\,\times 
\\[1.6ex]
\nonumber
&& \frac{1-x}{Q^{2}}\,
 \left[ 
     \left(\frac{Q^{2}}{-t}\right)^{b+1} +
     \left(\frac{Q^{2}}{-u}\right)^{b+1} +
    2\,t\,u\,
     \frac{
           \left( \left(\frac{Q^{2}}{-t}\right)^{\frac{b+1}{2}}       
                     -1 \right)    
               \left( \left(\frac{Q^{2}}{-u}\right)^{\frac{b+1}{2}}       
                     -1 \right)    
                                  }
      {(t+Q^{2})(u+Q^{2})} 
   \right] \, ,
\\[2.4ex]
\nonumber
\lefteqn{
 \frac{p_\mu \,p_{\nu}}{p\cdot q} \, 
\frac{d{\mathcal W}^{\mu \nu\,{(2)}}_{g}}{dt}
\, \left( \gamma^\ast + {\rm g}\rightarrow
    \overline{{\rm q}_L} + {\rm q}_R \right)   
 =}
\\[1.6ex]
\label{pmupnu}
&&\frac{e_q^2}{16}
   {\mathcal N}^{2}\,
\left( \frac{2\,\pi}{\alpha_s(\mu _{r} )}\right)^{13}\,
{\exp}\left[-\frac{4\,\pi}{\alpha_s(\mu _{r} )}\right] 
\,
\left( \frac{\mu _{r}^{2}}{Q^{2}}\right)^{b}\,\times 
\\[1.6ex]
\nonumber
&&
\frac{(1-x)^{2}}{Q^{2}}\,
t\,u\,
\left[
    \frac{\left(\frac{Q^{2}}{-t}\right)^{\frac{b+1}{2}}       
                     +\frac{u}{Q^{2}}\frac{x}{1-x}}{t+Q^{2}} 
   -\frac{\left(\frac{Q^{2}}{-u}\right)^{\frac{b+1}{2}}       
                     +\frac{t}{Q^{2}}\frac{x}{1-x}}{u+Q^{2}} 
\right]^{2} \, .
\end{eqnarray}
Here we have introduced the shorthand  
\begin{equation}
{\mathcal N} \equiv \sqrt{\frac{2}{3}}\,\pi^{2}\,d\ 
     2^{\,b}\,  \Gamma \left(\frac{b+1}{2}\right)\,
   \Gamma \left(\frac{b+3}{2}\right)\, .
\end{equation}
In Eqs.~(\ref{gmunu}) and (\ref{pmupnu}), the $t\leftrightarrow u$ symmetry
is manifest.  

%%%%%%%%%%%%%%%%%%%%%%%%%%%%%FIGURE  %%%%%%%%%%%%%%%%%%%%%%%%%%%%%%%%%%%
\begin{figure}
\begin{center}
  \epsfig{file=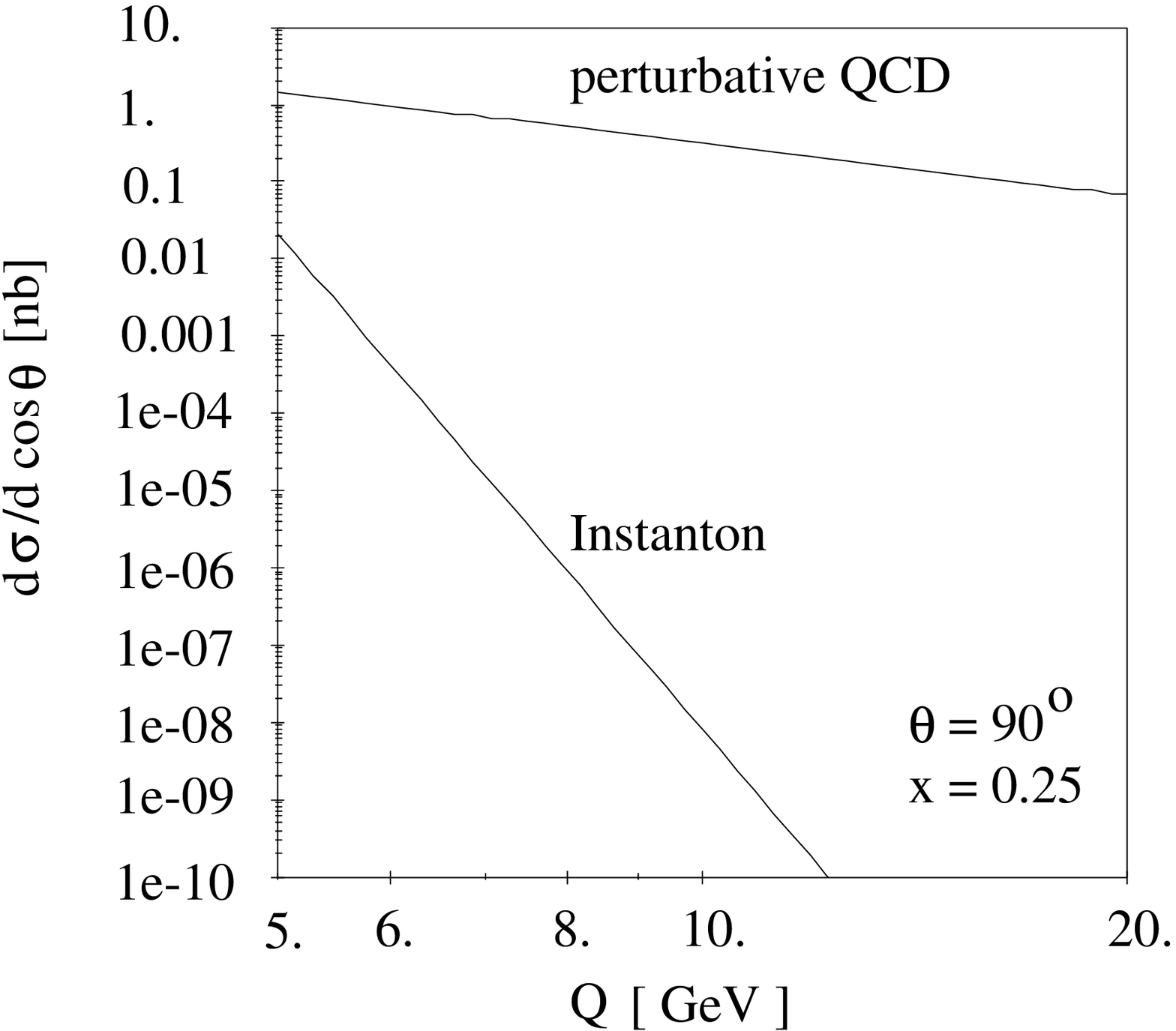,width=9cm}\vfill
  \hspace{8pt}
  \epsfig{file=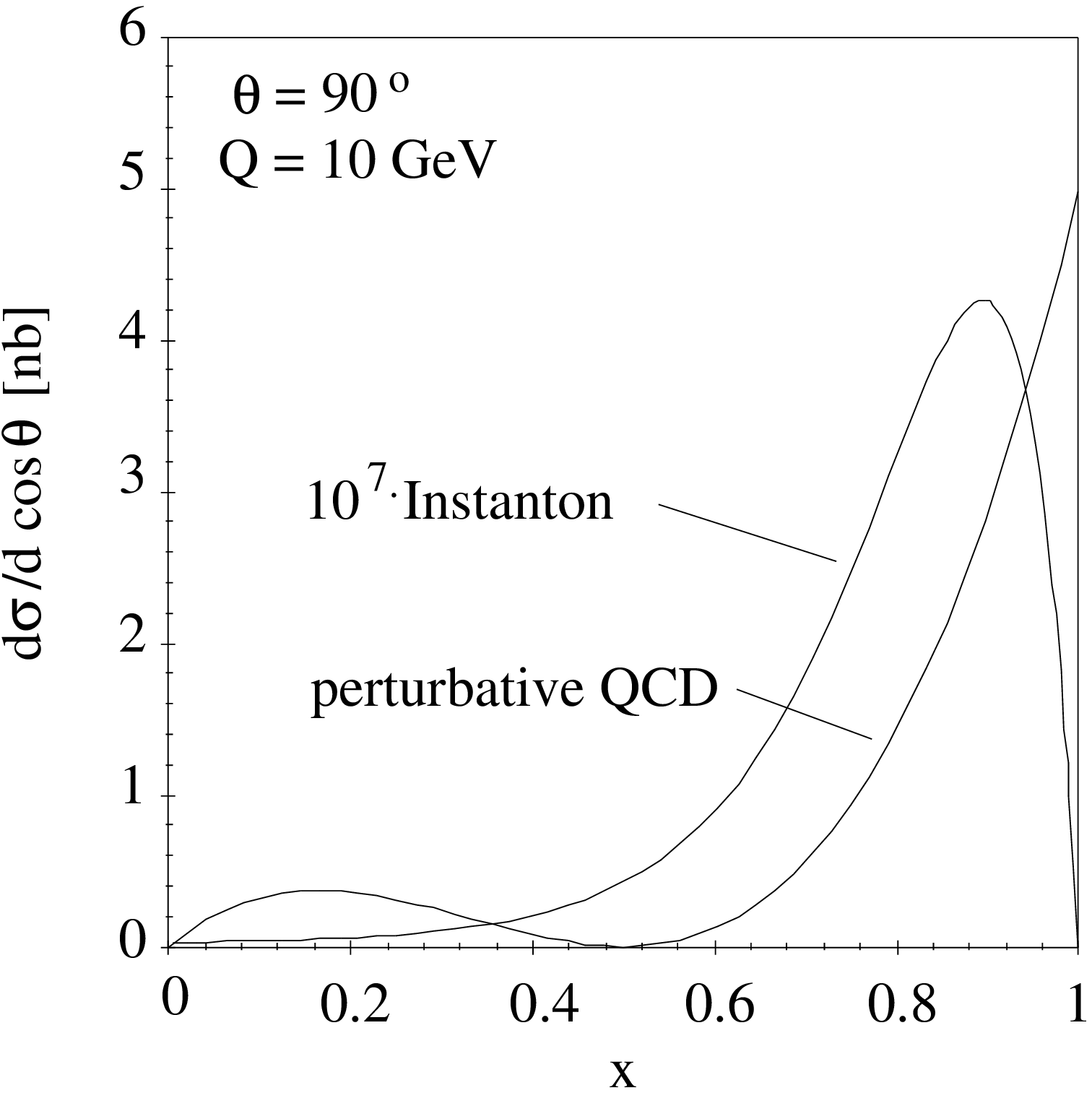,width=8cm}
\caption[dum]{\label{f4}
  Differential cross-sections, $d\sigma /d\cos\theta\ [\mbox{nb}]$, 
  of the $I$-induced {\it chirality-violating} process (\ref{instproc}) 
  and the perturbative {\it chirality-conserving} process
  (\ref{pertproc}), both for 
  fixed c.m. scattering angle, $\theta=90^{\,\circ}$, 
  $\Lambda=0.234$ GeV, $e_{q}=2/3$, and $\mu _{r}=Q$.
  Top: For fixed $x=0.25$, as function of $Q$ [GeV]. 
  Bottom:   
  For fixed $Q=10$ GeV, as function of $x$.}
\end{center}
\end{figure}
%%%%%%%%%%%%%%%%%%%%%%%%%%%%%%%%%%%%%%%%%%%%%%%%%%%%%%%%%%%%%%%%%%%%%% 

Upon inserting Eqs.~(\ref{gmunu}), (\ref{pmupnu}) into 
Eqs.~(\ref{diffcross}), (\ref{projF2}), and (\ref{projFL}), we see 
that the contribution of the $I$-induced process
(\ref{instproc}) to the differential cross-section $d\sigma /dt$
and the differential gluon structure functions, 
$d{\mathcal F}^{(2)}_{2\,g}/dt, d{\mathcal F}^{(2)}_{L\,{g}}/dt$, 
is well-behaved as long as we avoid the (collinear) singularities for
$t,u\rightarrow 0$. This is illustrated in Fig.~\ref{f4}, 
where we compare the differential cross-sections, $d\sigma /\cos\theta$, 
of both the $I$-induced process, Eq.~(\ref{instproc}), and the perturbative 
process, Eq.~(\ref{pertproc}), where  
\begin{equation}
\label{kt}
t=-\frac{Q^{2}}{2\,x}\,
\left( 1- \cos \theta 
\right) 
\, .
\end{equation}
We note that the renormalization-scale dependence of the $I$-induced
cross-section in Fig.~\ref{f4} is very small, due to the 
re\-nor\-ma\-li\-za\-tion-group im\-proved density (\ref{density}). 

Let us address at this point the important question concerning the
range of validity of the present calculation. Specifically, let us
examine the constraints emerging from the requirement of the 
dilute instanton gas approximation following Refs.~\cite{cdg,ag,as,svz}.
Along these lines one finds that instantons with size 
\begin{equation}
\label{rhoc}
\rho > \rho_{c}\simeq 1/(500\ {\rm MeV}) 
\end{equation}
are ill-defined semi-classically~\cite{svz}, 
corresponding to a breakdown of the dilute gas approximation.
On the other hand, using the form of our $\rho$ integral in 
Eqs.~(\ref{ampi2}), (\ref{V}), we may determine the average 
instanton size $\langle \rho\rangle$ contributing for a given
virtuality
\begin{equation}
\label{virt} 
{\mathcal Q}=\min \left( Q,
\sqrt{-t}=\frac{Q}{\sqrt{x}}\sin\frac{\theta}{2},
\sqrt{-u}=\frac{Q}{\sqrt{x}}\cos\frac{\theta}{2} 
\right)\, ,
\end{equation}
according to
\begin{equation}
\label{avrho}
\langle \rho\rangle \equiv 
\frac{\int\limits_{0}^{\infty}d\rho\, 
\rho\,\rho^{b+1}\,K_{1}(\rho\,{\mathcal Q})}
{\int\limits_{0}^{\infty}d\rho\, \rho^{b+1}\,K_{1}(\rho\,{\mathcal Q})}
\simeq 
\frac{b+3/2}{\mathcal Q}  \, .
\end{equation}
Hence, with Eq.~(\ref{rhoc}) and Eq.~(\ref{beff}), we find that
the virtuality $\mathcal Q$ should obey 
\begin{equation}
\label{qmin}
{\mathcal Q}\ (>)> (5-6)\ {\rm GeV}\, .
\end{equation}
In particular, our results in Fig.~\ref{f4} (top) for 
the $I$-induced differential
cross-section, $d\sigma /d\cos\theta$,  at $\theta=90^{\,\circ}$ and $x=0.25$, 
should be taken seriously only for $Q\ (= {\mathcal Q})> (5-6)$ GeV 
(since here $\sqrt{-t}=\sqrt{-u}=Q/\sqrt{2x}>Q$). 

Thus, like in the  perturbative case, {\it fixed-angle scattering
processes at high $Q^2$ are reliably calculable in (instanton) perturbation
theory} (at least in leading semi-classical approximation).

Next, we note that the contributions (\ref{gmunu}), (\ref{pmupnu}) of 
the $I$-induced exclusive process (\ref{instproc}) 
to the differential gluon structure functions are much more singular 
($\propto (-t, -u)^{-(b+1)}$) for  $t\to 0,u\to 0$  than the perturbative 
ones ($\propto (-t, -u)^{-1} $). This leads to a much stronger 
scheme dependence~\cite{mrs} than in the perturbative case.

Let us have a closer look at this feature. 
We regularize the collinear divergence of the $t$ integral 
along the same lines as in 
perturbation theory, i.e. we restrict the integration to the interval
$\{ -Q^{2}/x+\mu _{c}^{2},-\mu _{c}^{2}\}$. On account of 
Eqs.~(\ref{projF2}), (\ref{projFL}), we then obtain for   
the hard contributions of the $I$-induced exclusive process 
(\ref{instproc}) to the gluon structure functions,    
\begin{eqnarray}
\nonumber 
 {\mathcal F}_{2\,g}^{(\overline{{\rm q}_L}{\rm q}_R)}\,(x,Q^2;\mu _{c}^{2})
& =&\frac{e_q^2}{8}\, 
  {\mathcal N}^{2}\,
\left( \frac{2\,\pi}{\alpha_s(\mu _{r} )}\right)^{13}\,
{\exp}\left[-\frac{4\,\pi}{\alpha_s(\mu _{r} )}\right] 
\,\left( \frac{\mu _{r}^{2}}{\mu _{c}^{2}}\right)^{b}
\\[1.6ex]
\label{F2inst}
\mbox{}&& \times 
\frac{x\,(1-x)}{b}\,
\left[1+{\mathcal O}\left( \frac{\mu _{c}^{2}}{Q^{2}}\right)\right]
 \, ,
\\[2.4ex]
\nonumber
{\mathcal F}_{L\,g}^{(\overline{{\rm q}_L}{\rm q}_R)}\,(x,Q^2;\mu _{c}^{2})
& =&\frac{e_q^2}{2}\, 
   {\mathcal N}^{2}\,
\left( \frac{2\,\pi}{\alpha_s(\mu _{r} )}\right)^{13}\,
{\exp}\left[-\frac{4\,\pi}{\alpha_s(\mu _{r} )}\right] 
 \,\left( \frac{\mu _{r}^{2}}{\mu _{c}^{2}}\right)^{b}\,
\frac{\mu _{c}^{2}}{Q^{2}}\,
\\[1.6ex]
\label{FLinst}
\mbox{}&& \times \, 
\frac{x\,(1-x)^{2}}{b-1}\,
\left[1+{\mathcal O}\left(\frac{\mu _{c}^{2}}{Q^{2}}\right)\right]
 .
\end{eqnarray}

In the Bjorken limit, $Q^{2}/\mu _{c}^{2}\to \infty$, 
but with $\mu _{c}^{2}/\mu _{r}^{2}$ 
fixed, we find from Eqs.~(\ref{F2inst}) and (\ref{FLinst}), 
respectively, 
\begin{eqnarray}
\label{F2instbj}
\lefteqn{ 
\lim_{Q^{2}\to \infty}\,
 {\mathcal F}_{2\,g}^{(\overline{{\rm q}_L}{\rm q}_R)}(x,Q^2;\mu _{c}^{2})
 =}
\\[1.6ex]
\nonumber
&&
\frac{e_q^2}{8}\, 
  {\mathcal N}^{2}\,
\left( \frac{2\,\pi}{\alpha_s(\mu _{r} )}\right)^{13}\,
{\exp}\left[-\frac{4\,\pi}{\alpha_s(\mu _{r} )}\right] 
\,\left( \frac{\mu _{r}^{2}}{\mu _{c}^{2}}\right)^{b}\,
\frac{x\,(1-x)}{b}
 \, ,
\\[2.4ex]
\label{FLinstbj}
\lefteqn{
\lim_{Q^{2}\to \infty}\,
{\mathcal F}_{L\,g}^{(\overline{{\rm q}_L}{\rm q}_R)}(x,Q^2;\mu _{c}^{2})
 \equiv }
\\[1.6ex]
\nonumber
&&\lim_{Q^{2}\to \infty}\,\left[
{\mathcal F}_{2\,g}^{(\overline{{\rm q}_L}{\rm q}_R)}
(x,Q^2;\mu _{c}^{2})
-2\,x\,{\mathcal F}_{1\,g}^{(\overline{{\rm q}_L}{\rm q}_R)}
(x,Q^2;\mu _{c}^{2})\right]
= 0\, .
\end{eqnarray}
Hence, in this limit, the considered $I$-induced process gives a 
{\it scaling} contribution to ${\mathcal F}_{2\,g}$ and
the analogue of the Callan-Gross relation, 
${\mathcal F}_{2\,g}=2\,x\,{\mathcal F}_{1\,g}$, holds. 
In particular, this means that the {\it same} parton distribution
can absorb the infrared sensitivity of both structure functions, 
${\mathcal F}_{2\,g}^{(\overline{{\rm q}_L}{\rm q}_R)}$ 
and ${\mathcal F}_{1\,g}^{(\overline{{\rm q}_L}{\rm q}_R)}$.
This is one of the prerequisites of factorization~\cite{pert}.

\section{Conclusions and Outlook}

In this paper, we studied QCD-instanton induced processes in
deep-inelastic lepton-hadron scattering.
The purpose of the present work was to shed further light on the important 
questions around the IR-behaviour associated with the instanton size.
In order to eliminate unessential technical complications, we have  reduced
the realistic task of evaluating 
the $I$-induced cross-sections of the chirality violating multi-particle 
processes (illustrated in Fig.~\ref{f1})
\begin{equation}
  \gamma^{\ast}+{\rm g}\Rightarrow 
   \overline{{\rm u}_{L}}+{\rm u}_{R}\, 
   + \overline{{\rm d}_{L}}+{\rm d}_{R}\,
    +\overline{{\rm s}_{L}}+{\rm s}_{R}\,          
     +n_g\, {\rm g},
\end{equation}
to the detailed study of the {\it simplest} one,
without additional gluons and with just one massless flavour ($n_f=1$), 
\begin{equation}
  \gamma^{\ast}(q)+{\rm g}(p)\Rightarrow 
   \overline{{\rm q}_{L}}(k_{1})+{\rm
    q}_{R}(k_{2})\, .
\end{equation}
We have explicitly calculated the corresponding fixed angle cross-section 
and the contributions to the gluon structure functions 
within standard instanton perturbation theory in leading
semi-classical approximation (Sect.~3). To this approximation, 
{\it fixed-angle scattering processes at high $Q^2$ are reliably 
calculable}. In the Bjorken limit, the considered $I$-induced process gives a 
{\it scaling} contribution to ${\mathcal F}_{2\,g}$ and
the analogue of the Callan-Gross relation, 
${\mathcal F}_{2\,g}=2\,x\,{\mathcal F}_{1\,g}$, holds. 
 
All along we focused our main attention on the IR behaviour associated with
the instanton size. Gauge invariance was kept manifest along the 
calculation. 

As a central result of this paper and unlike Ref.~\cite{bb}, we found 
{\it no} IR divergencies associated with the integration over the 
instanton size $\rho$, which can even be perfomed analytically.  We have 
explicitly demonstrated  that the typical hard scale ${\mathcal Q}$ 
in deep-inelastic scattering provides a {\it dynamical} infrared cutoff for 
the instanton size, $\rho \lwig {\mathcal O}(1/{\mathcal Q})$. 
Thus, deep-inelastic scattering may indeed be viewed as a 
distinguished process for studying manifestations of QCD-instantons.   

In Ref.~\cite{bb}, the $I$-induced contribution to deep-inelastic scattering of
a virtual gluon from a real one~\cite{bbgg}, 
$g^\ast g \Rightarrow g^\ast g$,
served as a simplified r\^{o}le model for the splitting into a IR-finite 
contribution ($\rho \lwig 1/Q$) and an IR-divergent term (large $\rho$). 
As speculated  by one of the
authors~\cite{b}, the occurence of the IR-divergent term could well have
been due to the lacking gauge invariance of this model, associated with the
off-shellness of one of the initial gluons. In fact, one may enforce  
gauge invariance by replacing the instanton gauge field 
$A^{(I)}_\mu (x)$ describing the virtual gluon by the familiar 
gauge-invariant operator
\begin{equation}
G^{(I)}_\mu (x)=e^\alpha\, [x+\infty,x]\,
G^{(I)}_{\mu \,\alpha}(x)\,[x,x+\infty],
\end{equation}
where $G^{(I)}_{\mu \,\alpha}(x)$ is the instanton field-strength, 
and
\begin{equation}
[x,x+\infty]=P \exp\left\{ \ii\,g_{s} \int^\infty_0 d\lambda\,\, 
e\cdot A^{(I)}(x+\lambda e)
\right\}\, ,
\end{equation}
is a gauge factor ordered along the lightlike line in the direction
$e_\mu =(q_\mu +x p_\mu )/(2p\cdot q)$. In this case the IR-divergent term is, 
indeed, absent~\cite{b}.
Since our present calculation is manifestly gauge-invariant, the absence of
an IR divergent term fits well in line with these arguments.

A further main purpose of the present calculation was to provide a clean and
explicit discussion of most of the crucial steps involved in our subsequent
task~\cite{mrs} to calculate the {\it dominant} $I$-induced contributions
coming from final states with a large number of gluons (and three massless
flavours, say).

Let us close with some comments on the generalization
to the more realistic case with $n_g$ gluons in the final state, which is
entirely straightforward~\cite{mrs,q96}. Instead of Eq.~(\ref{ampi2}), 
the corresponding amplitude involves, in leading semi-classical approximation,
the additional factors from the $n_g$ gluons (c.f. Eq.~(\ref{lszgluon})),
\begin{eqnarray}
\label{ampi2g}
\lefteqn{
 {\mathcal T}_\mu ^{a\,a_1\ldots a_{n_g}}\,\left( \gamma^\ast + {\rm 
g}\rightarrow \overline{{\rm q}_L} + {\rm q}_R +n_g\,{\rm g}\right)   
 = }
\\[1.6ex]
\nonumber
&&
\ii\,e_q\,4\,\pi^2\,\left( 
\frac{\pi^3}{\alpha_s}\right)^{\frac{n_g+1}{2}}\,
\int dU
\int\limits_0^\infty d\rho\,d(\rho ,\mu _{r} )\,\rho^{2\,n_g} 
\\[1.6ex]
\nonumber
&&\times \,
{\rm tr}\left[ \sigma^{a}\,U\,\left[
\epsilon_g(p)\cdot p-\epsilon_g(p)\,\overline{p}
\right]\,U^{\dagger} \right]\,
\prod_{i=1}^{n_g}
{\rm tr}\left[ \sigma^{a_i}\,U\,\left[
\epsilon_g(p_i)\,\overline{p_i}-\epsilon_g(p_i)\cdot p_i
\right]\,U^{\dagger} \right]
\\[1.6ex]
\nonumber
&&\times 
\left\{
\left[ 
U \chi_R^\dagger (k_2 ) \epsilon 
\right]
\left[\epsilon V(q,k_1;\rho)\overline{\sigma}_\mu  
\chi_L(k_1)\, U^\dagger
\right] 
\right .
\\[1.6ex]
\nonumber
\mbox{}&&
- 
\left .
\left[ U \chi_R^\dagger (k_2) 
\sigma_\mu \overline{V}(q,k_2;\rho)\epsilon
\right]
\left[ 
\epsilon\, \chi_L(k_1)\, U^\dagger 
\right]
\right\}  ,
\end{eqnarray}
where the four-vector $V_\lambda$ is again given by Eq.~(\ref{V}). 
Besides the enhancement by a factor of $(\pi^3/\alpha_s)^{1/2}$, each
additional gluon gives rise to a factor of $\rho^2$ under 
the $I$-size integral. The IR-finiteness  of this integral is, however, 
not altered by the
presence of the additional overall factor of $\rho^{2\,n_g}$, on account of
the exponential cutoff $\propto \exp[ -\rho {\mathcal Q}]$ from the Bessel-K
function in the ``form-factors'' contained in $V_\lambda(q,k;\rho)$, 
Eq.~(\ref{V}). We also note, that the amplitude (\ref{ampi2g}) satisfies  
e.m. gauge invariance.

In analogy to electro-weak 
$(B+L)$-violation~\cite{m}, one expects~\cite{bb,rs} the sum of 
the final-state  gluon contributions to exponentiate, such that 
the total  $I$-induced $\gamma^\ast$g cross-section takes the form
(at large $Q^2$), 
\begin{eqnarray}
\label{exp}
\sigma^{{ (I)}}_{\gamma^\ast { g}}(x,Q^2) &\equiv&
\sum_{n_{ g}}\sigma^{{ (I)}}_{\gamma^\ast { g}\,n_{ g}}(x,Q^2)
\\[1.6ex]
\nonumber
&\sim& \int_x^1 dx^\prime\int\limits^{Q^2\frac{\xpr}{x}} 
\frac{dQ^{\prime 2}}{Q^{\prime 2}}\,\ldots\,
\frac{1}{Q^{\prime 2}}\,
\exp\left[-\frac{4\pi}{\alpha_s(Q^\prime)}\,F(x^\prime )\right]\, ,
\end{eqnarray}
where the so-called  ``holy-grail function''~\cite{m}
$F(\xpr )$ (normalized to F(1)=1) is expected to decrease towards 
smaller $\xpr$, which implies a dramatic growth of 
$\sigma^{{ (I)}}_{\gamma^\ast { g}}(x,Q^2)$ for decreasing $x$.

\section*{Appendix A}

Here we want to derive Eqs.~(\ref{onshellsi}) and 
(\ref{onshellsibar}) for the LSZ-amputated quark propagators. Let us 
first consider the Fourier transform of the
quark propagator (\ref{si}) which we write as 
\begin{equation}
S^{(I)}(x,-k)=\frac{1}{\sqrt{\Pi_x}}
\sum_{i=1}^3 s^{(i)} (x,-k) \, ,
\label{sumsi}
\end{equation}
where 
\begin{eqnarray}
\label{s1}
\lefteqn{
s^{(1)}(x,-k)
= }
\\[1.6ex]
\nonumber
&&
 2\left[ x- (-\ii\partial_k)\right]\,
(-\ii\overline{\partial}_k)\,(-\ii\partial_k)\,
\int \frac{d^4y}{(2\pi )^2} \frac{{\rm e}^{\ii k\cdot y}}
{\left( (x-y)^2\right)^2 \left( y^2+\rho^2\right)^{1/2}
\left( y^2\right)^{1/2}} \, ,
\\[2.4ex]
\label{s2}
\lefteqn{
s^{(2)}(x,-k)
=}
\\[1.6ex]
\nonumber
&&
 2\,\frac{\rho^2}{x^2}\left[ x- (-\ii\partial_k)\right]
U x (-\ii\overline{\partial}_k)\, U^\dagger
\int \frac{d^4y}{(2\pi )^2} \frac{{\rm e}^{\ii k\cdot y}}
{\left( (x-y)^2\right)^2 \left( y^2+\rho^2\right)^{1/2}
\left( y^2\right)^{1/2}} \, ,
\\[2.4ex]
\label{s3}
\lefteqn{
s^{(3)}(x,-k)
=} 
\\[1.6ex]
\nonumber
&&
\frac{\rho^2}{x^2}
\sigma_\mu U x\,
\left[ \overline{x}- (-\ii\overline{\partial}_k)\right]
\sigma_\mu (-\ii\overline{\partial}_k)\, U^\dagger
\int \frac{d^4y}{(2\pi )^2} \frac{{\rm e}^{\ii k\cdot y}}
{\left( x-y\right)^2 \left( y^2+\rho^2\right)^{3/2}
\left( y^2\right)^{1/2}} \, .
\end{eqnarray}
Our strategy to analyze the $k^2\to 0$ limit of Eqs.~(\ref{s1}-\ref{s3})
starts by partially evaluating the master integral,
\begin{equation}
I\,(-k;x,\rho ;\alpha , \beta , \gamma )
\equiv
\int \frac{d^4y}{(2\pi )^2} \frac{{\rm e}^{\ii k\cdot y}}
{\left( (x-y)^2\right)^\alpha \left( y^2+\rho^2\right)^\beta
\left( y^2\right)^\gamma} \, ,
\label{masterint}
\end{equation}
by means of the
Feynman parametrization (see e.g. \cite{yn}),
\begin{eqnarray}
\label{feynman}
\lefteqn{
\frac{1}{A^\alpha B^\beta C^\gamma}
=}
\\[1.6ex]
\nonumber
&& \frac{\Gamma (\alpha +\beta +\gamma ) }
{\Gamma (\alpha )\,\Gamma (\beta )\, \Gamma (\gamma )}\,
\int_0^1 da\,a \,
\int_0^1 db\,
\frac{(ab)^{\alpha -1} (a(1-b))^{\beta -1}\,(1-\alpha)^{\gamma 
-1}} {
\left[ ab\,A +a\,(1-b)\,B + (1-a)\,C\right]^{\alpha +\beta +\gamma}}
\, .
\end{eqnarray}
With the help of Eq.~(\ref{feynman}), 
it is possible to show that Eq.~(\ref{masterint}) can be expressed as
\begin{eqnarray}
\label{masterintfin}
\lefteqn{
I\,(-k;x,\rho ;\alpha , \beta , \gamma )
=}
\\[1.6ex]
\nonumber
&& \frac{2^{1-(\alpha +\beta +\gamma )}}{ \Gamma (\alpha )\,
\Gamma (\beta )\, \Gamma (\gamma )}\,
\int_0^1 da\,a^{\alpha +\beta  -1}\,(1-a)^{\gamma -1}
\int_0^1 db\,b^{\alpha  -1}\,(1-b)^{\beta  -1}\,
{\rm e}^{\ii\,k\cdot x\,ab} 
\\[1.6ex]
\nonumber
&&
\times \,\left( 
\frac{\sqrt{x^2 ab(1-ab)+\rho^2 a(1-b)}}{\sqrt{k^2}}
\right)^{2-(\alpha +\beta +\gamma )}
\\[1.6ex]
\nonumber
&&
\times \,K_{2-(\alpha +\beta +\gamma )} 
\left(
\sqrt{k^2}\sqrt{x^2 ab(1-ab)+\rho^2\, a(1-b)}
\right) .
\end{eqnarray}

Next we insert Eq.~(\ref{masterintfin}) into  
Eqs.~(\ref{s1})-(\ref{s3}), perform the various derivatives, and expand
the integrand with respect to $k^2\to 0$. Finally, the remaining Feynman 
parameter integrations are done. After this procedure we find:
\vfill\eject 
\begin{eqnarray}
\lim_{k^{2}\to 0}{s^{(1)}} 
(x,-k)(-\ii\,\overline{k}) &=& 
-{\rm e}^{\ii\,k\cdot x}\, ,
\\[1.6ex]
\lim_{k^{2}\to 0}{s^{(2)}}
(x,-k)(-\ii\,\overline{k}) &=&- \frac{1}{2}\frac{\rho^2}{x^2}
\frac{\left[ U x \overline{k} U^\dagger
\right]}{k\cdot x} 
\left[ {\rm e}^{\ii\,k\cdot x} -
\frac{\ii}{k\cdot x} 
\left(1-{\rm e}^{\ii\,k\cdot x}\right)\right]
 ,
\\[1.6ex]
\lim_{k^{2}\to 0}{s^{(3)}}
(x,-k)(-\ii\,\overline{k}) &=& \frac{1}{2}\frac{\rho^2}{x^2}
\frac{\left[ U x\overline{k} U^\dagger
\right]}{k\cdot x}
\left[ 1 -
\frac{\ii}{k\cdot x}
\left(1-{\rm e}^{\ii\,k\cdot x}\right)\right]
 .
\end{eqnarray}
Thus, on account of Eq.~(\ref{sumsi}), the on-shell 
residuum of the quark propagator (\ref{si}) is given by 
Eq.~(\ref{onshellsi}). A similar reasoning leads to 
Eq.~(\ref{onshellsibar}) for the residuum of the quark propagator 
(\ref{sibar}).

\section*{Appendix B}

Our task is to derive Eqs.~(\ref{tvertex}), (\ref{uvertex}), corresponding 
to the $\gamma^{\ast}$- quark vertex ${\mathcal V}_\mu ^{(t,u)}$ in the 
leading-order $I$-induced amplitude. We will concentrate on 
the derivation of Eq.~(\ref{tvertex}), since the derivation of 
Eq.~(\ref{uvertex}) is completely analogous. 
 
Let us recall the definition of ${\mathcal V}_\mu ^{(t)}$, but now with
indices written explicitly, 
\begin{eqnarray}
{\mathcal V}^{(t)}_{\mu \  m \lambda}\,(q,-k) =
\int d^4x\,{\rm e}^{-\ii\,q\cdot x}\ 
\overline{\phi}^{\dot\alpha}_l (x)\,
\overline{\sigma}_{\mu \,\dot\alpha\alpha}\,
\lim_{k^2\to 0}\, 
{{S}^{(I)\,\alpha\dot\beta\ l}}_m\,(x,-k)\,
\left(-\ii\,\overline{k}_{\dot\beta\lambda}\right)
 .
\label{phivertex1}
\end{eqnarray}
Inserting Eqs.~(\ref{phibar}) and (\ref{onshellsi}) 
into Eq.~(\ref{phivertex1}) we obtain for the vertex, 
\begin{eqnarray}
{\mathcal V}^{(t)}_{\mu \ m \lambda}\,(q,-k) &=&
-\frac{\rho^{3/2}}{\pi}\, 
\int d^4x\,{\rm e}^{-\ii\,( q-k)\cdot x}\,
\frac{1}{\left( x^2+\rho^2\right)^2}
\label{phivertex2}
\\
&\times &
\epsilon_{\gamma\delta}\,
\left[ x\,\overline{\sigma}_\mu \right]^\delta_{\ \lambda}\,
\left[
\left( U^\dagger\right)^\gamma_{\ m}
+\frac{1}{2}\,\frac{\rho^2}{x^2}\,
\frac{\left[ x\,\overline{k}\,U^\dagger\right]^\gamma_{\ m}}
{k\cdot x}\,
\left( 1-{\rm e}^{-\ii\,k\cdot x}\right) 
\right]\, .
\nonumber 
\end{eqnarray}
The matrix structure in Eq.~(\ref{phivertex2}) can be simplified 
using  
\begin{equation}
\epsilon_{\gamma\delta}\,
\left[ x\,\overline{\sigma}_\mu \right]^\delta_{\ \lambda}\,
\left[ x\,\overline{k}\,U^\dagger\right]^\gamma_{\ m}
= x^2\,\epsilon_{\gamma\delta}\,\left( U^\dagger\right)^\gamma_{\ m}\,
\left[ k\,\overline{\sigma}_\mu \right]^\delta_{\ \lambda}\, ,
\end{equation}
which follows from the transposition rules of the $\sigma$-matrices. 
Thus Eq.~(\ref{phivertex2}) can be rewritten as
\begin{eqnarray}
{\mathcal V}^{(t)}_{\mu \ m \lambda}\,(q,-k) &=&
-\frac{\rho^{3/2}}{\pi}\, 
\int d^4x\,{\rm e}^{-\ii\,( q-k)\cdot x}\,
\frac{1}{\left( x^2+\rho^2\right)^2}
\label{phivertex3}
\\
&\times &
\epsilon_{\gamma\delta}\,\left( U^\dagger\right)^\gamma_{\ m}\,
\left[
\left[ x\,\overline{\sigma}_\mu \right]^\delta_{\ \lambda}
+\frac{1}{2}\,\rho^2\,
\left[ k\,\overline{\sigma}_\mu \right]^\delta_{\ \lambda}\,
\frac{1}{k\cdot x}\,
\left( 1-{\rm e}^{-\ii\,k\cdot x}\right) 
\right]\, .
\nonumber
\end{eqnarray} 
The remaining $d^4x$ integration in Eq.~(\ref{phivertex3}) can be
done with the help of the following formulae ($k^2=0$ is always understood), 
\begin{eqnarray}
\label{int1}
\lefteqn{
\int d^4x\,{\rm e}^{-\ii\,(q-k)\cdot x}\, 
\frac{x}{\left( x^2+\rho^2\right)^2} 
=}
\\[1.6ex]
\nonumber
&&
-2\,\pi^2\,\ii\,\frac{q-k}{\left( q-k\right)^2}\,
\rho\,\sqrt{\left( q-k\right)^2}\,
K_1\left(\rho\,\sqrt{\left( q-k\right)^2}\right)\, ,
\\[2.4ex]
\label{int2}
\lefteqn{
\int d^4x\,{\rm e}^{-\ii\,(q-k)\cdot x}\, 
\frac{1}{\left( x^2+\rho^2\right)^2\,\left( k\cdot x\right)} 
= }
\\[1.6ex]
\nonumber
&&
2\,\pi^2\,\ii\,
\frac{1}{q\cdot k}\,\frac{1}{\rho^2}\,
\rho\,\sqrt{\left( q-k\right)^2}\,
K_1\left(\rho\,\sqrt{\left( q-k\right)^2}\right)\, ,
\\[2.4ex]
\label{int3}
\lefteqn{
\int d^4x\,{\rm e}^{-\ii\,q\cdot x}\, 
\frac{1}{\left( x^2+\rho^2\right)^2\,\left( k\cdot x\right)} 
= }
\\[1.6ex]
\nonumber
&&
2\,\pi^2\,\ii\,
\frac{1}{q\cdot k}\,\frac{1}{\rho^2}\,
\rho\,\sqrt{q^2}\,
K_1\left(\rho\,\sqrt{q^2}\right)\, .
\end{eqnarray}
By means of these basic integrals, we obtain finally  for the
vertex, 
\begin{eqnarray}
\lefteqn{
{\mathcal V}^{(t)}_{\mu \ m \lambda}\,(q,-k) 
=
2\,\pi\,\ii\,\rho^{3/2}\,
\epsilon_{\gamma\delta}\,\left( U^\dagger\right)^\gamma_{\ m} } 
\label{phivertex4}
\\
&&\times 
\Biggl\{ 
\frac{\left[ \left(q-k\right) 
\,\overline{\sigma}_\mu \right]^\delta_{\ \lambda}}
{(q-k)^2}\,
\rho\,\sqrt{\left( q-k\right)^2}\,
K_1\left(\rho\,\sqrt{\left( q-k\right)^2}\right)
\nonumber
\\
\mbox{}&&-
\frac{1}{2}\,
\frac{\left[ k\,\overline{\sigma}_\mu \right]^\delta_{\ \lambda}}
{q\cdot k}\, \left[
\rho\,\sqrt{\left( q-k\right)^2}\,
K_1\left(\rho\,\sqrt{\left( q-k\right)^2}\right)
-\rho\,\sqrt{q^2}\,
K_1\left(\rho\,\sqrt{q^2}\right)
\right]
\Biggr\}\, ,
\nonumber 
\end{eqnarray}
in accordance with Eq.~(\ref{tvertex}). 

\vspace{10pt}
\section*{Acknowledgements}
We would like to acknowledge helpful discussions with 
V. Braun and V. Rubakov.


\vspace{10pt}
\begin{thebibliography}{99}

\bibitem{bpst}
A. Belavin, A. Polyakov, A. Schwarz and Yu. Tyupkin, 
\Journal{\PLB}{59}{85}{1975}.
\bibitem{th} 
G. `t Hooft, \Journal{\PRL}{37}{8}{1976};
\Journal{\PRD}{14}{3432}{1976}; \Journal{\PRD}{18}{2199}{1978}
(Erratum).
\bibitem{r} A. Ringwald, \Journal{\NPB}{330}{1}{1990}; 
\hfill\break
O. Espinosa, \Journal{\NPB}{343}{310}{1990}.
\bibitem{m}
M. Mattis, \Journal{\PRC}{214}{159}{1992};
\hfill\break
P. Tinyakov,
\Journal{\IJMPA}{8}{1823}{1993}; 
\hfill\break
R. Guida, K. Konishi and N. Magnoli,
\Journal{\IJMPA}{9}{795}{1994}. 
\bibitem{bb} 
I. Balitsky and V. Braun, \Journal{\PLB}{314}{237}{1993}. 
\bibitem{rs}
A. Ringwald and F. Schrempp, DESY 94-197, hep-ph/9411217, in:
{\it Quarks `94}, Proc. VIIIth Int. Seminar, Vladimir, Russia, 
May 11-18, 1994, eds. D. Grigoriev {\it et al.}, pp. 170-193.
\bibitem{grs}
M. Gibbs, A. Ringwald and F. Schrempp, DESY 95-119,\\ hep-ph/9506392,
in: {\it Proc. Workshop on Deep-Inelastic Scattering and QCD},
Paris, France, April 24-28, 1995, eds. J.-F. Laporte and Y. Sirois,
pp. 341-344.
\bibitem{rs1}
A. Ringwald and F. Schrempp, DESY 96-125, hep-ph/9607238, to appear 
in: {\it Proc. Workshop DIS96 on Deep-Inelastic Scattering and Related
Phenomena}, Rome, Italy, April 15-19, 1996.
\bibitem{ggmrs}
M. Gibbs, T. Greenshaw, D. Milstead, A. Ringwald and F. Schrempp, 
``Search Strategies for Instanton-Induced Processes at HERA'', to appear in:
Proc. {\it Future Physics at HERA}, 1996.
\bibitem{H1}
H1 Collaboration, S. Aid {\it et al.}, DESY 96-122, hep-ex/9607010,
submitted to Nucl. Phys.
\bibitem{bbgg}
I. Balitsky and V. Braun, \Journal{\PRD} {47} {1879}{1993}. 
\bibitem{as}
T. Appelquist and R. Shankar,
\Journal{\PRD}{18}{2952}{1978}.
\bibitem{early}
For further related discussions, see:\\
N. Andrei and D. Gross,
\Journal{\PRD}{18}{468}{1978};
\hfill\break
L. Baulieu, J. Ellis, M. Gaillard and W. Zakrzewski, 
\Journal{\PLB}{77}{290}{1978}; \Journal{\PLB}{81}{41}{1979};
\hfill\break
V. Novikov, M. Shifman, A. Vainshtein and V. Zakharov,
\Journal{\NPB}{174}{378}{1980};
\hfill\break
M. Dubovikov and A. Smilga, \Journal{\NPB}{185}{109}{1981}.
\bibitem{mrs}
S. Moch, A. Ringwald and F. Schrempp, to be published.
\bibitem{q96}
A. Ringwald and F. Schrempp, DESY 96-203, to appear in:
{\it Quarks `96}, Proc. IXth Int. Seminar, Yaroslavl, Russia, May 5-11, 1996.
\bibitem{f}
R. Field, {\it Applications of Perturbative QCD}, (Addison-Wesley,
New York, 1989).
\bibitem{brown}
L. Brown, R. Carlitz, D. Creamer and C. Lee,
\Journal{\PRD}{17}{1583}{1978}.
\bibitem{ber}
C. Bernard, \Journal{\PRD}{19}{3013}{1979}.
\bibitem{abc}
A. Vainshtein, V. Zakharov, V. Novikov and M. Shifman,
\Journal{\SPU}{25}{195}{1982}.
\bibitem{morretal}
T. Morris, D. Ross and C. Sachrajda, 
\Journal{\NPB}{255}{115}{1985}.
\bibitem{bbb}
I. Balitsky, M. Beneke and V. Braun,
\Journal{\PLB}{318}{371}{1993}.
\bibitem{yn} F. Yndur{\'a}in, {\it The Theory of Quark and Gluon
Interactions}, (Springer-Verlag Berlin Heidelberg, 1993).
\bibitem{cdg}
C. Callan, R. Dashen and D. Gross, 
\Journal{\PRD}{17}{2717}{1978}.
\bibitem{ag}
N. Andrei and D. Gross, 
\Journal{\PRD}{18}{468}{1978}.
\bibitem{svz}
M. Shifman, A. Vainshtein and V. Zakharov, 
\Journal{\NPB}{165}{45}{1980}.
\bibitem{pert}
J. Collins, in: {\it Perturbative Quantum Chromodynamics}, 
edited by A. Mueller, (World Scientific, Singapore, 1989).
\bibitem{b}
I. Balitsky, Pennsylvania State Univers. preprint, PSU-TH-146 (94/05), 
hep-ph/9405335, in: {\it Continuous advances in QCD}, Proc.,  
Minneapolis, USA, Feb. 18-20, 1994, ed. A. Smilga, pp.167-194. 



\end{thebibliography}
\end{document}